\documentstyle[twoside,fleqn,espcrc2,epsf]{article}

\newcommand{\del}{{\bf D}}

\newcommand{\delv}{{\bf \del}}

\newcommand{\delsq}{D^{(2)}}

\newcommand{\lag}{{\cal L}}

\newcommand{\mbare}{\mbox{$M_0$}}

\newcommand{\nl}{\nonumber \\}

\newcommand{\Bv}{{\bf B}}

\newcommand{\sigmav}{\mbox{\boldmath$\sigma$}}

\newcommand{\gammav}{\mbox{\boldmath$\gamma$}}

\newcommand{\be}{\begin{equation}}
\newcommand{\ee}{\end{equation}}

\newcommand{\order}{{\cal O}}

\newcommand{\AmS}{{\protect\the\textfont2
  A\kern-.1667em\lower.5ex\hbox{M}\kern-.125emS}}

\title{ Perturbative Matching of the NRQCD Heavy-Light Axial Current }

\author{
        Junko Shigemitsu
        \address{Physics Department, 
        The Ohio State University, 
        Columbus, OH 43210, USA.}
        and 
        Colin J. Morningstar 
        \address{Physics Department, University of California at San Diego,
 La Jolla, CA 92093, USA.}
        \thanks{Talk presented by J.Shigemitsu for the GLOK Collaboration 
at LATTICE '97, Edinburgh, July 1997. }
                           }
       
\begin{document}

\begin{abstract}

A one-loop matching calculation between Lattice NRQCD and full QCD 
for the heavy-light axial current is described and 
 the effects of renormalization on $f_B$ are discussed.

\end{abstract}

\maketitle

\section{Introduction}

\noindent
An important ingredient in lattice studies of B meson decays, is the matching 
between currents in full continuum QCD and current operators of the 
lattice theory being simulated.  The GLOK collaboration, for instance, 
uses NRQCD to simulate b-quarks together with 
tadpole improved clover light quarks \cite{glok}.  One  goal is to calculate 
the pseudoscalar meson decay constant $f_{PS}$,
\be  \label{deffb}
   \langle \, 0 \,| \, A_{\mu} \,|\,PS\, \rangle = p_{\mu} f_{PS} . 
\ee
  In general, several operators of 
the effective theory are associated with a given full QCD current 
operator and one has, for instance, 
\be 
{\langle  \, A_0 \, \rangle}_{QCD}
= \sum_j C_j \langle \, J^{(j)} \, \rangle_{NRQCD}
\ee
Our task is to identify the relevant operators $J^{(j)}$ and 
calculate the matching coefficients $C_j$ to a given order in $\alpha$ 
and $1/M$. We present here results of a matching calculation through 
\order($\alpha/M$) \cite{pert}
 and discuss their implications for $f_B$.

\vspace{.4in}

\section{ The Lattice Action and Current Operators}

\noindent
The lattice NRQCD heavy quark action density used in the perturbative
 calculations is 

 \begin{eqnarray} \label{nrqcdact}
 \lag_{NRQCD} &= & \overline{\psi} \psi -  
 \overline{\psi}
\left(1 \!-\!\frac{a \delta H}{2}\right)
 \left(1\!-\!\frac{aH_0}{2n}\right)^{n} \nl
 &&  U^\dagger_4  
  \left(1\!-\!\frac{aH_0}{2n}\right)^{n}
\left(1\!-\!\frac{a\delta H}{2}\right) \psi
 \end{eqnarray}
with 
 \be
 H_0 = - {\delsq\over2\mbare} \, , \qquad 
\delta H  = - c_B \frac{g}{2\mbare}\,\sigmav\cdot\Bv
 \ee
$c_B = 1$ at tree level. The one-loop contribution to $c_B$ 
 is a higher order (\order($\alpha^2/M$) ) effect in the 
matching calculation and hence can be ignored here.
For the light quarks we use the clover action.  A few comments will be made 
later, on perturbative calculations with Wilson light quarks.

\vspace{.1in}
\noindent
Heavy-light currents in full QCD have the form  $\bar{q} \Gamma h$.  The 
four component Dirac spinor for the heavy quark, $h$, is related to the 
two component NRQCD heavy quark (heavy anti-quark)
fields, $\psi$ ($\tilde{\psi}$), via an inverse Foldy-Wouthuysen 
transformation.
Through \order($ 1/M$) one has,
\be
h = U^{-1}_{FW} \Psi_{FW} = [ 1 - \frac{1}{2 M}(\gammav \cdot \delv) ] \,
 \left(\begin{array}{c} 
                                        \psi  \\
                                      \tilde{\psi}
                       \end{array}  \right)  
\ee
Hence at tree-level,
\begin{eqnarray} \label{jtree}
J &=& J^{(0)} + J^{(1)}  \nonumber \\
  &=& \bar{q} \Gamma Q - \frac{1}{2 M} \bar{q} \Gamma (\gammav
\cdot \delv) Q
\end{eqnarray}
$Q = \frac{1}{2}(1 + \gamma_0) \Psi_{FW} $.\\
At one-loop one needs to include, for $ \Gamma = \gamma_5 \gamma_0$ and
 massless light quarks, a third  operator
\be
  J^{(2)} = 
 \frac{1}{2 M} (\delv \bar{q} \cdot \gammav )\Gamma  Q
\ee
Furthermore,  on the lattice there is a discretization correction to 
$J^{(0)}$,
\be
J^{(0)} \rightarrow J^{(0)}_{imp} \equiv J^{(0)} + C_A J^{(disc)}
\ee
with
\be
J^{(disc)} = a\,(\delv \bar{q} \cdot \gammav )\Gamma  Q
\ee
$C_A$ is fixed by requiring that $\langle J^{(0)}_{imp} \rangle$ 
projects only onto matrix elements that exist in the continuum theory.  
In perturbation theory $C_A$ starts out \order($\alpha$).  $J^{(disc)}$ 
is the analogue of the discretization correction $a\,\partial_\mu P$ 
($P$ = pseudoscalar density) to 
the axial current in light physics emphasized by the Alpha collaboration. 
Recent studies have shown that this term has a non-negligible 
effect on $f_\pi$ around $\beta = 6.0$.  In heavy-light physics the 
improvement term $J^{(disc)}$ to $J^{(0)}$ also leads to a significant 
correction to $f_B$.

\section{ Matching and One-Loop Correction Terms}

\begin{figure}[t]
\epsfxsize=9.0cm
\centerline{\epsfbox{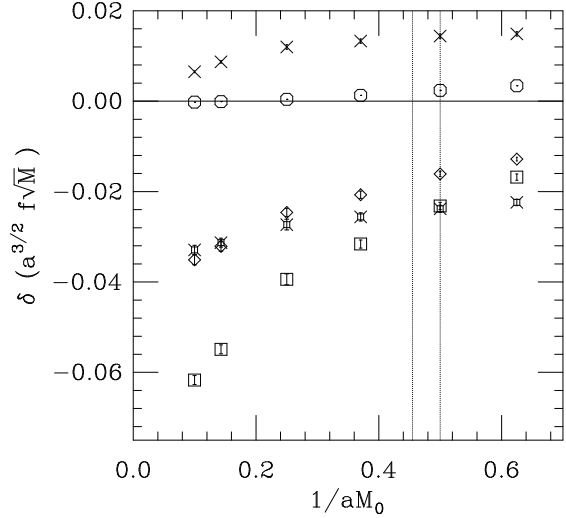}}
\caption{ One-loop Corrections to $a^{3/2}f_{PS}\protect\sqrt{M_{PS}}$ from 
$\protect\rho_0 \, \protect\langle J^{(0)} \protect\rangle$ (Diamonds) , 
$\protect\rho_1 \, \protect\langle J^{(1)} \protect\rangle$ (Circles) , 
$\protect\rho_2 \, \protect\langle J^{(2)} \protect\rangle$ (Crosses) , 
$\protect\rho_d \, \protect\langle J^{(disc)} \protect\rangle$ (Fancy Squares).
Simple squares denote the total one-loop correction. 
 Vertical lines mark the physical B.
 }

\end{figure}

\noindent
In order to determine the matching coefficients $C_j$ we consider scattering 
of a heavy quark by the heavy-light current ($\bar{q} \Gamma h$ or the 
$J^{(i)}$'s)  into  
a light quark.  The calculation is carried out in both full QCD and 
in lattice NRQCD and one requires that the scattering amplitudes in the 
two theories agree through \order($\alpha/M$).  For the continuum QCD 
calculation we use NDR in the $\overline{MS}$ scheme. A gluon mass $\lambda$ 
is introduced at intermediate stages as an IR regulator.  IR divergences 
cancel between the continuum and lattice contributions to the $C_j$'s.  
After taking mixing among the $J^{(i)}$'s into account one ends up 
with the following one-loop expression,
\begin{eqnarray} \label{ajlat}
{\langle  \, A_0 \, \rangle}_{QCD}
&=& \sum_j C_j \langle \, J^{(j)} \, \rangle   \nl
&=&(1 + \alpha \, \rho_0) \left[ \langle J^{(0)}\rangle + C_A
\langle J^{(disc)}\rangle \right]  \nl
&+&(1 + \alpha \, \rho_1) \,\langle J^{(1)}\rangle \nl
&+&\qquad  \alpha \, \rho_2 \;\, \langle J^{(2)}\rangle 
\end{eqnarray}
The matrix elements on the RHS are evaluated in the effective theory NRQCD. 
Writing $C_A = \alpha \, \rho_d$, one sees that there are four one-loop 
correction terms to $\langle A_0 \rangle$ and hence also to $f_{PS}$
\be
\alpha\,\rho_0 \, \langle J^{(0)} \rangle \quad , \quad 
\alpha\,\rho_1 \, \langle J^{(1)} \rangle \quad , \quad 
\alpha\,\rho_2 \, \langle J^{(2)} \rangle 
\ee
and
\be
\alpha\,\rho_d \, \langle J^{(disc)} \rangle 
\ee
In Fig.1 we show contributions from these terms to $a^{3/2} f_{PS} 
\sqrt{M_{PS}}$.  Simulation results for the matrix elements were obtained 
on $\beta = 6.0$ quenched configurations and results are for $ \kappa = 
\kappa_{strange}$ and $\alpha = \alpha_P(1/a)$. 
One should note the relative importance of the discretization correction 
term $\alpha \, \rho_d \,\langle J^{(disc)} \rangle$.  At the B-meson, due 
to cancellation among the other one-loop corrections, this is the dominant 
one-loop term which, as we will see in the next section, leads to a 
$\sim 12$\% reduction of $f_B$ relative to the tree-level result.  
  Both $\alpha \, \rho_d \langle J^{(disc)} \rangle$ and 
$\alpha \, \rho_0 \langle J^{(0)} \rangle$ survive into the static limit.  
The other two terms are \order($\alpha/M$) and vanish in that limit.

\vspace{.1in}
\noindent
The results presented sofar are for clover light quarks.  For Wilson 
light quarks some modifications are necessary. 
In particular, 
  if one tries to calculate 
$C_A = \alpha \rho_d$ with Wilson light quarks,
 one finds an uncancelled logarithmic  IR divergence 
$\frac{- 2 \alpha}{ 3 \pi}\,a\, ln(a\lambda)$.  This divergence can only be 
removed by including clover contributions to the calculation. 
Hence, it is not possible to include $\langle J^{(disc)} \rangle$ in 
a consistent way within perturbation theory if one is working with Wilson 
light quarks. 
 ( this is in addition to the fact that since Wilson light 
quarks have \order($a$) errors one normally would not include \order($a \, 
\alpha$) corrections).
 A figure similar to Fig.1 for Wilson quarks would only have 
contributions from $\rho_i \langle J^{(i)} \rangle$, i = 0,1,2.  Although the 
numbers change slightly upon going from clover to Wilson, the qualitative 
feature of cancellation among the three one-loop corrections around the 
B meson still holds.  As a result the difference between tree-level and 
one-loop corrected $f_B$ is smaller for Wilson than for clover.  On the 
other hand there will be large scaling violations due to the omission 
of the sizeable $J^{(disc)}$ improvement term.

\section{ Results for $f_B$}

\noindent
Fig. 2 shows preliminary GLOK collaboration results for, \\
$a^{3/2} \Phi \equiv $ [ $a^{3/2} f_{PS} \sqrt{M_{PS}}$ ] with $ln (aM)$ 
in $C_0$ and $C_1$ set to $ln (aM_b) \approx ln(2.2)$ for all $M_0$. This 
ensures a smooth $aM \rightarrow \infty$ limit and contact with 
HQET scaling formulas while preserving the correct $f_B$ for the 
physical B meson.  
%
%
 $q^*$ for the coupling $\alpha_P(q^*)$ is not known yet for this calculation 
and we show results for $aq^* = \pi$ and $aq^* = 1$.  
Using the $\rho$-meson mass to fix $a^{-1}$ one finds,

\begin{figure}[t]
\epsfxsize=9.0cm
\centerline{\epsfbox{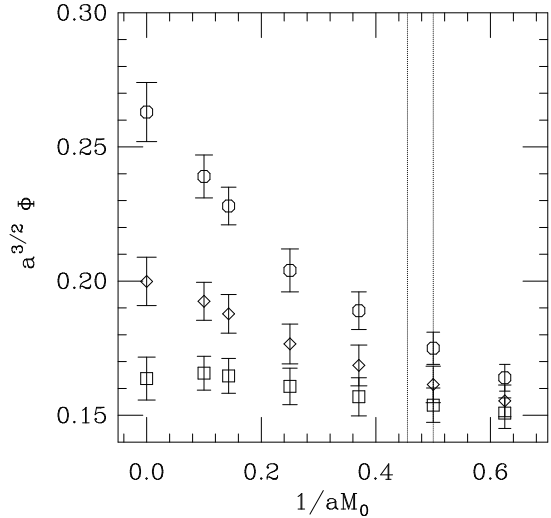}}
\caption{  Heavy-Light Decay Constants 
at $\kappa_s$.
 Circles : tree-level ; Diamonds : one-loop matching with $aq^* = \pi$ ;
Squares : one-loop matching with $aq^* = 1$. Vertical lines mark 
the physical B.
 }
\end{figure}

\[f_{B_s} =
\cases{ 
0.207\,(9)\; GeV  \qquad tree-level  \cr
0.189\,(8)\; GeV  \qquad aq^* = \pi  \cr
0.178\,(8)\; GeV  \qquad aq^* = 1  \cr } \]

\noindent
Only statistical errors are shown. 
After extrapolating to $\kappa_{light}$,

\[f_{B} =
\cases{ 
0.168\,(13)\; GeV  \qquad tree-level  \cr
0.153\,(12)\; GeV  \qquad aq^* = \pi  \cr
0.144\,(11)\; GeV  \qquad aq^* = 1  \cr } \]

\noindent
In both cases, one-loop corrections are a 9\% (14\%) effect for $aq^* = 
\pi (1)$.  Taking an average, the preliminary estimate for quenched $f_B$
at $\beta = 6.0$ is,

$$f_B = 0.149 \, (12)(^{+22}_{\;-\,5})(9) \, GeV  $$

\noindent
The second error comes from scale uncertainties and the third from 
higher order relativistic and perturbative corrections.


\begin{thebibliography}{99}


\bibitem {glok}
See contributions by Arifa Ali Khan and Joachim Hein to these proceedings.


\bibitem {pert}
C.Morningstar and J.Shigemitsu, in preparation. 
J.Shigemitsu,  hep-lat/9705017.

\end{thebibliography}
\end{document}